\documentclass[11pt,a4paper,12pt]{article}
\usepackage{epsfig}
\usepackage{float}
\usepackage{graphicx}
\usepackage{amssymb}
\begin{document}
\thispagestyle{empty}

\newcommand{\etal}  {{\it{et al.}}}  
\def\Journal#1#2#3#4{{#1} {\bf #2}, #3 (#4)}
\def\PRD{Phys.\ Rev.\ D}
\def\NIMA{Nucl.\ Instrum.\ Methods A}
\def\PRL{Phys.\ Rev.\ Lett.\ }
\def\PLB{Phys.\ Lett.\ B}
\def\EPJ{Eur.\ Phys.\ J}
\def\IEEETNS{IEEE Trans.\ Nucl.\ Sci.\ }
\def\CPCD{Comput.\ Phys.\ Commun.\ }



{\Large\bf
\begin{center}
Dark photons in the Dalitz-like decay of a scalar
\end{center}
}

\begin{center}
\large{ G.A. Kozlov  }
\end{center}
\begin{center}
 { Bogolyubov Laboratory of Theoretical Physics\\
 Joint Institute for Nuclear Research,\\
 Joliot Curie st., 6, Dubna, Moscow region, 141980 Russia }
\end{center}


 \begin{abstract}
 \noindent
 {The couplings of the Standard Model sector to the scale invariant degrees of freedom can open the possibility to study dark photons (DP).  The Dalitz-like decay of the (Higgs-like) scalar boson into a single photon and DP is studied. 
The interaction between DP and quarks is mediated by the derivative of the scalar field - the dilaton, the virtual (fictitious) state. The mass of the dilaton does not enter the final solutions. 
The limits are set on the DP mass, the mixing strength between the standard photon and DP. 
}


\end {abstract}




\bigskip

{\bf I. Introduction.-}  It is known the interest to  the theories in which the Standard Model (SM) is strongly coupled to a conformal sector of particles (see the refs. 4-22 in [1]) and to the unparticle stuff [2]. In theories where an exact conformal symmetry is spontaneously broken, the low energy effective theory contains a massless scalar boson, the dilaton. The nature of the dilaton couplings is governed by the requirement that conformal symmetry be realized non-linearly. At the loop level, the dilaton has potentially enhanced couplings to photons and gluons compared to those of the SM Higgs. However, these couplings are model dependent. In contrast to the SM Higgs, the dilaton couples to gluons even before running any SM particles in the loop, through the trace anomaly.

We investigate the conformal anomaly of the type scalar-vector-vector ($SVV$) containing a scale (dilatation) current $K_{\mu}$ and two vector currents which are related to each other, and this $SVV$ anomaly reflects the violation of scale invariance in SM by quantum effects. 
The certain gauge models may admit the additional $U^{\prime}(1)$ gauge group associated with new  vector bosons  which can have small masses or will be almost massless (for a recent review see [3] and the references therein).
In these models the extended group $SU(2)_{L}\times U(1)_{Y}\times U^{\prime}(1)_{B}$ does appear, where  index $B$ in  $U^{\prime} (1)_{B}$ may be associated with a hidden vector sector containing an extra photon $\bar\gamma$. The standard photon $\gamma$ may oscillate into  $\bar\gamma$  where the latter would be a short-lived dark photon (DP) decaying to invisible neutrino-antineutrino pair, $\bar\gamma\rightarrow \bar\nu\,\nu$ or to light lepton-antilepton pair $l\bar l$. 
We develop the model in which the coupling of the SM-scale anomaly to electromagnetic sector can lead to production of a SM photon and DP. 
The coupling to the scale invariant degrees of freedom (d.o.f.) - the scalar dilaton  - is most important operator in our analysis.  
Once the conformal invariance is broken, the dilaton gets the vacuum expectation value (vev) $f$. The breaking of conformal invariance at the scale $\Lambda_{CFT} = 4\,\pi\,f$ triggers electroweak symmetry breaking (EWSB) at the scale $\Lambda_{EW} = 4\,\pi\,v < \Lambda_{CFT}$, where $v =246$ GeV is the vev of the Higgs boson. The scales $f$ and $v$ are different except for the Higgs boson ($f = v$). If the approximate conformal symmetry is broken at $\Lambda_{CFT}$, the low energy spectrum of composite states may contain a light dilaton, a light Higgs doublet or both.
The dilaton operator triggers the breaking of  $SU(2)_{L}\times U(1)_{Y}$ gauge invarince through  the dilaton mass operator. 

The hidden sector containing DP  can be explored in collider experiments at high energies. The dark photon mixes with the standard photon via the kinetic term $\sim \epsilon\,F_{\mu\nu}\,B^{\mu\nu}$ with $\epsilon$ being the  mixing strength; $F_{\mu\nu}$ and $B_{\mu\nu}$ are the strength tensors of electromagnetic field $A_{\mu}$ and $B_{\mu}$ field of DP, respectively. The basic object is
 the Dalitz-like decay  $S\rightarrow \gamma\,\bar\gamma$, where $S$ should either be the SM Higgs boson $H$ or the scalar dilaton, where the latter  is the conformal portal to dark sector.
So, when a scalar boson is detected in the Higgs sector, it is important to determine whether it is $H$ or another scalar, e.g., the dilaton.


In this paper, we consider the scenario, where  DP  is within the reach of the LHC energy $\sqrt s \sim O(10~TeV)$. 
Recently, the DP and resonant monophoton signatures in Higgs boson decays at the LHC have been studied in [4].
The mixing strength $\epsilon$ is predicted in various models with the values in the range $10^{-12} - 10^{-2}$. In the low energy experiments the  values of $\epsilon$ in the window $10^{-7} - 10^{-3}$ have been probed (see the refs. in [5]). 
If no excess events are found, the obtained results can be used to impose bounds on the $\gamma-\bar\gamma$ mixing strength $\epsilon$ as a function of DP mass.
The production of $S$ is due to gluon-gluon fusion followed by the (heavy) quarks in the loop with final states $\gamma$ and $\bar\gamma$.
In SM, the violation of conformal symmetry is understood through the non-vanishing  divergence of the dilatation current,  $\partial_{\mu} K^{\mu}$
$$\partial_{\mu}K^{\mu} = \theta^{\mu}_{\mu} = \frac{\beta (g)}{2\,g}\, G^{a}_{\mu\nu}\,G^{\mu\nu\,a} + \sum_{q} m_{q} [1 + \gamma_{m} (g^{2}) ]\bar{q}\,q.$$
Here, $\theta^{\mu}_{\mu}$ is the trace of the energy-momentum tensor, $\beta (g)$ is the standard  beta-function with the coupling constant $g$, $G^{a}_{\mu\nu}$ is the strength of gluon tensor. The quark state $q$ is accompanied by the mass $m_{q}$ and $\gamma_{m}(g^{2})$ stands for the anomalous dimension of the mass operator $\bar q q$. If the conformal invariance is breaking, the scalar color-singlet state $S (p)$ with the mass $m_{S}$,  the momentum $p_{\mu}$ and the decay constant $f_{S}$ can be produced when $K_{\mu}$ acts on the vacuum
$$ \langle 0\vert K^{\mu} (x)\vert S(p)\rangle = i\,p^{\mu}\,f_{S}\,e^{-i\,p\,x},\,\,\, \langle 0\vert \theta^{\mu}_{\mu}(x)\vert S(p)\rangle = m^{2}_{S}\,f_{S}\,e^{-i\,p\,x}, $$ 
where $\vert 0\rangle$ is the vacuum state corresponding to spontaneously broken dilatation symmetry.
The coupling of the triangle scale anomaly to $\gamma\,\bar\gamma$ final state is given by the effective interaction
$$L_{S\,\gamma\,\bar\gamma} = \epsilon\,g_{S\,\gamma\,\bar\gamma}\,F_{\mu\nu}\,B^{\mu\nu}\,S, $$
where the origin of $S$ is undestood through $\theta^{\mu}_{\mu}$ due to conformal anomaly; $g_{S\,\gamma\,\bar\gamma}$ is the coupling of the effective interaction 
related  to $f_{S}^{-1}$ and can be fixed from the experiment.

Because of the scale invariance, the operator relevant to $\bar\gamma$ may carry the features of an unparticle  stuff [2] with the scaling dimension in the infra-red (IR), $d_{IR} = 1+ \delta$, $\delta <1$. In this case, the energy spectrum $\omega_{\gamma\bar\gamma} = d\,\Gamma (S\rightarrow\gamma\bar\gamma)/dE_{\gamma}$ of the photon (with energy $E_{\gamma}$)  has a continuous distribution spreading from zero to $m_{S}/2$. As $\delta \rightarrow 0^{+}$ (from above) 
$\omega_{\gamma\bar\gamma}$ becomes more peaked at $m_{S}/2$. The $\bar\gamma$ with $\delta = 0$ is a massless standard photon. 
Note that the unitarity constraint lower bound on Conformal field theory (CFT) vector operator dimension $d_{IR} \geq 3$ [6] 
does not use here because the operator of $\bar\gamma$ is both not gauge invariant and not primary one. 
The decay of $\bar\gamma$ into SM particles is controlled by the relation between the mass gap $m_{\bar\gamma}$ of $\bar\gamma$ and the production threshold $m_{1} + m_{2}$ with the masses $m_{i}$ $(i=1,2)$ of the decay products. If $m_{\bar\gamma} > m_{1} + m_{2}$ there is enough phase space that $\bar\gamma$ can decay into SM particles, otherwise DP may be associated with dark matter as a stable object.


In this paper DP is considered in the framework of the gauge dipole field model which exhibits IR singularities. In Abelian Higgs model the breaking of the gauge symmetry implies the dipole singularity of the type $\delta^{\prime} (p^{2})$  in two-point Wightman function (TPWF), e.g., for the scalar  or the gauge fields,  satisfying the equations of motion of 4th order [7-12]. 
Other classes of models exhibiting  
a  $\delta^{\prime} (p^{2})$ singularity are the CFT models  
[13,14].

In Sec. II the couplings and constraints of the DP through the observable $\epsilon$ are found. Sec. III is devoted to the Lagrangian which defines the Higgs-dilaton Abelian gauge model, and to the resulting equations of motion. In Sec. IV we obtain TPWF and the propagator of DP. In Sec. V the physical and virtual states are considered through asymptotic expressions for physical observables. In the concluding section the results are summarized.

{\bf II. Couplings and constraints.-} Conformal field theory coupled to SM can be a candidate to describe the hidden sector, containing, in particular, DP. The ultraviolet (UV) coupling of an operator $O_{UV}$ of dimension $d_{UV}$ to a SM operator $O_{SM}$ of dimension $d$ at the UV scale $M$ (UV messenger) has the form
\begin{equation}
\label{e2}
\frac{1}{M^{d-4}} \,O_{SM} \frac{1}{M^{d_{UV}}}\,O_{UV}.
\end{equation}
No masses are allowed in the Lagrangian of the effective theory containing (\ref{e2}). All masses  can be generated dynamically in IR.
We consider the hidden sector which is formed itself when the dilaton field $\bar\sigma (x)$ is coupled to a $U^{\prime}(1)$ gauge theory. The coupling of $\bar\sigma$ to DP sector in UV is
$$\frac{1}{M^{d_{UV}-2}} \,{\vert\bar\sigma\vert }^{2}\,O_{UV},$$
which flows in the IR to coupling of the Higgs boson $H$ to DP operator $O_{IR}$ of dimension $d_{IR}$ 
$$const\,\frac{\Lambda^{d_{UV} - d_{IR}}}{M^{d_{UV}-2}} \,{\vert H\vert }^{2}\,O_{IR}$$
when the scale invariance is almost breaking. The scale of the strong coupling sector $\Lambda$  is proportional to $f$.  Once $H$ acquires $v$, theory becomes nonconformal below the scale $\tilde\Lambda$, where [15]
$$\tilde\Lambda^{4- d_{IR}} = \frac{\Lambda^{d_{UV} - d_{IR}}}{M^{d_{UV}-2}} \,v^{2}.$$
Below $\tilde\Lambda$ the hidden sector becomes a standard particle sector. For a typical energy $\sqrt {s}$ of a collider experiment $\tilde\Lambda < \sqrt {s} <\Lambda$, which leads to the energy constraint  to search for DP physics
$$s^{2- d_{IR}/2} > \left (\frac{\Lambda}{M}\right )^{d_{UV} - d_{IR}} \,M^{2 - d_{IR}} \,v^{2}.$$
Based on the operator form $(\ref{e2})$ the mixing strength $\epsilon$, as an observable, is 
\begin{equation}
\label{e8}
\epsilon = \left (\frac {\sqrt {s}}{M}\right )^{2(d-4)}\, \left (\frac {\sqrt {s}}{\Lambda}\right )^{2d_{IR}}\, \left (\frac {\Lambda}{M}\right )^{2d_{UV}}. 
\end{equation}
The effect of DP sector on observables has no the dependence on $d_{UV}$, $d_{IR}$ and $\Lambda$, and is restricted by 
\begin{equation}
\label{e9}
\epsilon < \frac{s^{d}}{\left (v^{2}\,M^{d-2}\right )^{2}}.
\end{equation}
It is clear from (\ref {e8}) that signals of new physics with DP increase with energy, 
and would be seen if the values of the parameter $M$ are not too large. If the deviation from the SM is detected at the level of order $3\%$, the DP would be visible at the LHC ($\sqrt{s}\sim$ O(10 TeV)) as long as $v < M < 1000$ TeV with  $d=4$.
In the decay $S\rightarrow\gamma\bar\gamma$ the gauge invariant operator structure is
 $O_{SM}\,O_{IR}\sim \epsilon\bar q\,\gamma_{\mu}\,q\,S\, B^{\mu}\,M^{-1}$
and the relevant energy scale  is the mass of the heavy quark $q$ in the loop. 
Since the new effects beyond the SM is expected on the scale $M > v \sim O(0.3 ~ TeV)$, we find the upper limit on the mixing strength $\epsilon < 3\cdot 10^{-2}$
in the case of the top-quark in the loop  ($d=4$).
If the mixing $\epsilon$ is going to zero, the only decay of the SM Higgs boson into two photons would be appropriate. 

The amplitude of the decay $\bar\gamma$ into $\nu\bar\nu $, is
$$Am (\bar\gamma\rightarrow \nu\bar\nu) = \frac{1}{2}\,f_{\nu}\,\bar\nu \left (g_{V_{\nu}}\,\gamma_{\beta} + g_{A_{\nu}}\,\gamma_{\beta}\,\gamma_{5} \right)\,\nu\,\bar\gamma_{\beta}, $$
where  $f^{2}_{\nu} = 4\,\sqrt {2} \,G\,m^{2}$, $G\sim 10 ^{-5}~GeV^{-2}$ is the weak coupling strength.
Since there are no final state interactions in $\bar\gamma\rightarrow \nu\bar\nu $ decay, one has its partial width in the form 
$\Gamma (\bar\gamma\rightarrow \nu\bar\nu) =(\bar\alpha/3)\,\epsilon^{2}\cdot m$, where 
$\epsilon^{2} =  {G\,g^{2}_{\nu}\,m^{2}}/{(\sqrt{2}\,\pi\,\bar\alpha)}$, $m$ is the mass of DP, $\bar\alpha$ is an electromagnetic gauge coupling, 
and $g^{2}_{V_{\nu}} = g^{2}_{A_{\nu}} \equiv g^{2}_{\nu}$ is taken into account. 
Using the  restriction on $\epsilon < 3\cdot 10^{-2}$ above mentioned, one can find the upper limit on the DP mass $m < 3.3 $ GeV.  
 The calculation of the electromagnetic neutrino formfactor (EM$\nu$F) [16], applied to the  process $\bar\gamma\rightarrow\nu\bar\nu$, gives the DP mass in the form
$$ m \simeq m_{\mu} \left [ 3\,\sqrt{2}\,\pi\,\bar\alpha^{-1}\,{\sum_{l:e,\mu}\left (\ln \frac{\Lambda^{2}_{\nu}}{m_{l}^{2}} -\frac{1}{6}\right)}^{-1}\right ]^{\frac{1}{2}}.$$
Here, $m_{l}$ is the mass of the charged lepton in the loop ($l: e, \mu$), $\Lambda_{\nu}$ is the cut off scale in the logarithmically divergent integral of EM$\nu$F. Using $\Lambda_{\nu}$ at the scale of the $Z$-boson mass, the DP mass $m = 0.83$ GeV is equated for the case of an electron and muon loops in EM$\nu$F, and as the result, the mixing strength is $\epsilon = 7.6\cdot 10^{-3}$.  In case of the electron loop only, the DP is very light (with the mass $m = 4.5$ MeV) and short-lived state which can decay also into $e^{+}e^{-}$ pair with a lifetime bounded by $\tau > 10^{-16}$ sec.
The results obtained in this section for  $\epsilon$ and $m$ are consistent with the constraints in $\epsilon$ vs $m$ plane based on the low energy experiments data (see the refs. in [5]). 

{\bf III. Model.-} The  Lagrangian density (LD) of  the Higgs-dilaton Abelian gauge model  where the  field $B_{\mu}$ mixes with  $A_{\mu}$ is
\begin{equation}
\label{e10}
L_{\epsilon} = -\frac{1}{2}\,\epsilon\,F_{\mu\nu}\,B^{\mu\nu} - \xi (\partial_{\mu} A^{\mu})(\partial_{\nu} B^{\nu}) + \bar q (i\hat\partial - m_{q} - g\,\hat A)\,q - I^{\mu} (B_{\mu} - \partial_{\mu}\sigma),
\end{equation}
where $\xi =\epsilon\bar\xi$,
$\bar\xi$ is a real number, $g$ is dimensioneless coupling constant,  
$F_{\mu\nu} = \partial_{\mu} A_{\nu} - \partial_{\nu} A_{\mu}$, 
$B_{\mu\nu} = \partial_{\mu} B_{\nu} - \partial_{\nu} B_{\mu}$;
 $m_{q}$ is the quark mass, and $I_{\mu}$ is an auxiliary field.  The subcanonical scalar field $\sigma (x)$ with zero dimension in mass units is the primary dilaton field (the grandfather  potential) which provides a control over UV and IR divergences.
LD (\ref{e10}) is invariant under the restricted gauge transformations of the second kind
\begin{equation}
\label{e11}
A_{\mu}\rightarrow A_{\mu} +\partial_{\mu}\alpha, \,\,\, B_{\mu}\rightarrow B_{\mu} +\partial_{\mu}\alpha,\,\, \sigma\rightarrow \sigma + \alpha,\,\, q\,\rightarrow q \,e^{i\,g\,\alpha},\,\, I_{\mu}\rightarrow I_{\mu}, 
\end{equation}
where $\alpha (x)$ obeys the equation $\Box\alpha (x) = 0$ ($ \Box \equiv \partial_{\mu} \,\partial^{\mu}$).  
If the SM is the part of CFT the Higgs couplings to massless gauge bosons with $q$-fields in the loop is replaced by corresponding dilaton couplings  $(2\,m^{2}_{q}/v^{2})H^{+}\,H (x) \rightarrow m^{2}_{q}\,\sigma^{2}(x)$ [17].
The parametrization of the $\sigma$ couplings to quarks that are relevant for collider physics is
\begin{equation}
\label{e122}
L_{\sigma} = -{\sigma}\,\sum_{q^{\prime}} (m_{q^{\prime}} + x_{\sigma}\,y_{q^{\prime}}\,v)\,q^{\prime}\,\bar q^{\prime},
\end{equation}
where  $x_{\sigma} = m^{2}_{\sigma} /f^{2} <1$ parametrizes the size of deviations from exact scale  invariance; $m_{\sigma}$ is the mass of the dilaton; $y_{q}$ are 9 additional contributions to the Yukawa couplings. 
The dilaton $\sigma$ serves as  a conformal compensator which under gauge transformation shifts as a Goldstone boson, $\sigma\rightarrow \sigma + \alpha$.
The invariance of LD (\ref{e10}) is broken because of  (\ref{e122}) and a small mass parameter $m$ of DP if they should incorporate in (\ref{e10}). 
If SM is embedded in the conformal sector we assume that  $q^{\prime}$ are those quark d.o.f. which obey the condition 
\begin{equation}
\label{e1222}
\sum_{light} b_{i} + \sum_{heavy} b_{i} = 0, 
\end{equation}
where $i$ carries either QCD or electroweak (EW) features of the coefficients $b_{i}$ of corresponding $\beta$-functions.
The sum in (\ref{e1222}) is splitted over all colored particles into sums over light and heavy states in the mass scale separated by  $m_{\sigma}$. 
Hence the only quarks $q^{\prime}$ in (\ref{e122}) lighter than that of the dilaton are included in the coefficients of corresponding $\beta$-function. 
In particlular, $b^{light}_{QCD} = - c_{G} = -11 + 2n_{light}/3$, where the number of light quarks is either  $n_{light} = 5$ if $m_{\sigma} < m_{t}$, or $n_{light} = 6$ if $m_{\sigma} > m_{t}$ for the top quark mass $m_{t}$ [17]. In EW sector one has $b^{light}_{EW} = -80/9$ if $m_{\sigma} < 2 m_{W}$, or $b^{light}_{EW} = -35/9$ if $2m_{W} < m_{\sigma} < 2 m_{t}$, or $b^{light}_{EW} = -17/3$ if $2 m_{t} < m_{\sigma}$, where $m_{W}$ is the mass of the $W$-boson [18].
Because of conformal condition (\ref{e1222}) the dilaton contribution to the decay $S\rightarrow\gamma\,\bar\gamma$ corresponds to including the states lighter than the dilaton in the $\beta$-function coefficients.

We consider the scale invariant sector of the theory, which contains the  fields $H$ and $\bar\sigma$ 
\begin{equation}
\label{e13}
L_{SI} (x) = \frac{1}{2}\left [\left (\partial_{\mu} H\right )^{2}  + \left (\partial_{\mu} \bar\sigma\right )^{2}\right ] - \frac{\lambda}{4}\left (H^2 -\beta^2 \bar\sigma^{2}\right )^{2} - \frac{\eta}{4} \left (\bar\sigma^{2} - f^{2}\right )^{2}.  
\end{equation}
The LD (\ref{e13}) is invariant under dilatation transformations $x\rightarrow x^{\prime} = c\,x$ $(c > 0)$, $\phi (x) \rightarrow \phi^{\prime} (x^{\prime}) = c^{-d_{\phi}}\,\phi (x)$, where $d_{\phi}$ is the scaling dimension of $\phi$, where $\phi: H,\,\bar\sigma$;  $\beta = \langle H\rangle /\langle\bar\sigma\rangle = v/f$. The conformal symmetry is breaking explicitly by the mass term $\sqrt {\eta /2}\,f$ which is supposed to be small compared to $f$. If $\eta =0$ we deal with the dilaton having the flat direction.  
The LD of the model becomes
\begin{equation}
\label{e14}
L = L_{\epsilon} + L_{\sigma} -\frac{1}{2}\,m ^{2}\,B_{\mu}\,B^{\mu} + L_{SI}. 
\end{equation}

The equations of motion are
$$(i\hat\partial - m_{q} -
g\hat A)\,q =0,\,\,\,   \sigma\sum_{q^{\prime}} (m_{q^{\prime}}  + x_{\sigma}\,y_{q^{\prime}}\,v)\,q^{\prime} = 0,$$
$$\xi\Box (\partial\cdot A) = f\left [\lambda\beta^{2}(H^{2} - \bar\sigma^{2}) - \eta \bar\sigma ^{2}  +  2 m^{2}_{\sigma} -  \left (1 - \frac{m^{2}}{f^{2}}\right)\Box \right ]\bar\sigma  - \sum_{q^{\prime}} (m_{q^{\prime}} + x_{\sigma}y_{q^{\prime}}v)\bar q^{\prime}q^{\prime},$$
\begin{equation}
\label{e17}
\Box B_{\mu} = \frac{1}{\epsilon} Y_{\mu} + \left (1-\bar\xi\right ) \partial_{\mu}(\partial\cdot B),\,\,\, Y_{\mu} =g\bar q\,\gamma_{\mu}\,q,
\end{equation}
\begin{equation}
\label{e18}
B_{\mu} = \frac{\epsilon}{m^{2}} \left [\Box A_{\mu} - \frac{1}{\epsilon}\,I_{\mu} -\left (1-\bar\xi\right ) \partial_{\mu}(\partial\cdot A)\right ].
\end{equation}
The equation of motion for $q$ implies current conservation $\partial_{\mu} Y^{\mu} = 0$. 
Then taking into account the equations  (\ref{e17}) and  $B_{\mu}(x) = \partial_{\mu}\sigma (x)$, one approaches the dipole equation for the Hermitian virtual dilaton  field $\bar\sigma (x)$  
\begin{equation}
\label{e21}
 \Box^{2}\bar\sigma (x) =  0,\,\,\,\, f\neq 0.
\end{equation}
Eq. (\ref{e21}) means the conformal symmetry is restored because $\Box \bar\sigma (x) \neq 0$.
Since the TPWF of dilaton field does not admit the Kallen-Lehmann representation, $\bar\sigma (x)$  is the quantum field defined in the space with an indefinite metric.


{\bf IV. TPWF and propagator.-} 
It is known that the space with an indefinite metric is useful to the formalization of the idea of virtual (or potential) states. 
For this, the Hilbert space of physical state vectors is decomposed into two parts: one is for the real (physical) particles and another one contains the virtual state vectors. We shall consider this subject in the next section. 

Let us consider the TPWF for $\bar\sigma (x)$ in the form
\begin{equation}
\label{e22}
\omega (x) = \langle\Omega, \bar\sigma (x)\, \bar\sigma (0)\Omega\rangle.
\end{equation}
 The vacuum vector $\Omega$ in (\ref{e22}) is defined as the vector satisfying $\bar\sigma^{(-)} (x)\Omega = 0$, $\langle\Omega,\Omega\rangle = 1$, where $\bar\sigma (x)$ is decomposed into negative-frequency (annihilation) and positive-frequency (creation) parts: $\bar\sigma (x) = \bar\sigma^{(-)} (x) + \bar\sigma^{(+)} (x)$, $\bar\sigma^{(+)} (x) = [\bar\sigma^{(-)} (x)]^{+}$.  
The function $\omega (x)$ (\ref{e22}) should be Lorentz-invariant, and the equation 
\begin{equation}
\label{e23}
\Box^{2} \omega (x) = 0
\end{equation}
is evident from (\ref{e21}). The general solution of (\ref{e23}) is given by the following expansion (see [9] and the refs. therein)
\begin{equation}
\label{e24}
\omega (x) = b_{1}\,\,ln\frac{l^{2}}{-x^{2}_{\mu} +i\,\varepsilon\,x^{0}} + b_{2}\,\frac{1}{x^{2}_{\mu} - i\,\varepsilon\,x^{0}} + b_{3}
\end{equation}
which is the distribution on the space $S^{\prime}(\Re ^{4})$ of temperate generalized function on $\Re ^{4}$. The space $S^{\prime}(\Re ^{4})$ is conjugate to the (complex) space  $S (\Re ^{4})$ of the test functions on $\Re ^{4}$.  
The coefficients $b_{1}$ and  $b_{2}$ in (\ref{e24}) will be defined later, while $b_{3}$ is an arbitrary constant.
The parameter $l$ in (\ref{e24}), having the dimension in units of length,  breaks the scale invariance under dilatation transformation.
The TPWF (\ref{e22}) is the homogeneous generalized function of the zeroth order with the dilatation properties 
\begin{equation}
\label{e244}
\omega (\rho x) = \omega (x) - 1/(8\,\pi)^{2}\,\ln\rho,\,\,\,\rho > 0. 
\end{equation} 

The Fourier transformation (FT) of the first term in (\ref{e24}) is given by  
$$\int 2\,\pi\,\theta (p^{0})\,\delta^{\prime}(p^{2}, \hat M^{2})\,e^ {-ipx} d_{4}p, $$ where $\hat M = (2/l)\, e^ {1/2-\gamma}$, $\gamma $ is the Euler's constant. Here, we run into the non-defined product $ \theta (p^{0})\,\delta^{\prime}(p^{2}, \hat M^{2})$ of generalized functions (IR divergence). This product is well-defined distribution only on the space $S(\Re_{4})$ of complex Schwartz test functions $u(p)$. In particular,
$$ \delta^{\prime}(p^{2}, \hat M^{2}) = \frac{1}{16} \Box_{p}^{2} \left [ \theta (p^{2})\,\ln \frac{p^{2}}{\hat M^{2}}\right ]. $$
and would have the form as $\delta^{\prime} (p^{2})$ for the same IR reasons only on the space
$S_{0}(\Re_{4})$, where the test functions $u(p)$ from $S(\Re_{4})$ are zero at $p = 0$. 
The Hermitian form  $\langle\Omega, \sigma (\tilde f)\, \sigma (\tilde g)\Omega\rangle$ ($\tilde f, \tilde g \in \mathbb S (\Re _{4}))$ on $S (\Re _{4})$  does not defined as the positive one. The reason of the latter is followed from 
(\ref{e244}) as well as from 

$$\int 2\,\pi\,\,\tilde f (p)\,\theta (p^{0})\,\delta^{\prime}(p^{2})\,\tilde g (p)\, d_{4} p = \int_{\Gamma_{0}^{+}} \, \frac{1}{2\,n\,p} \,\left ( -n\,\partial + \frac{1}{n\,p}\right )\,\tilde f (p) \,\tilde g (p) \, \frac{d^{3} p}{2p^{0}},$$
where $n$ is the fixed unit time-like vector ($n^{2} =1 $) in the Minkovsky space $\mathcal {S} (\mathbb {M})$ from $V^{+} = \{p \in\mathbb {M}, p^{2} > 0, p^{0} > 0\}$; $\Gamma_{0}^{+} = \{p \in \mathbb {M}, p^{2} = 0, p^{0} > 0\}$, $n\partial = n_{\mu}\,\partial/{\partial p_{\mu}}$. Therefore, the functional $\theta (p^{0})\,\delta^{\prime} (p^{2})$ is defined as to the distribution from $S^{\prime}(\Re ^{4})$ through the first term in  (\ref{e24}), where $l$ is the parameter of IR regularization.
For $n = (1, \vec {0})$ one has
$$- \frac{\partial}{\partial m_{\sigma}^{2}} \int 2\,\pi\,\theta (p^{0})\,\delta(p^{2} - m_{\sigma}^{2})\,u (p)\, d_{4} p = - \frac{\partial}{\partial m_{\sigma}^{2}} \int \frac{d_{3} p}{2\,E} \,u (E, \vec {p}). $$
One can verified that
$(p^{2})^{2}\,\delta^{\prime} (p^{2}, \hat M^{2}) = 0, \,\,\, p^{2}\,\delta^{\prime} (p^{2}, \hat M^{2}) = - \delta (p^{2}).$
The presence of $\delta^{\prime} (p^{2}, \hat M^{2})$ in FT of 
$\omega (x)$ is a consequence of the nonunitarity of translations ($\delta^{\prime} (p^{2}, \hat M^{2})$ is not a measure). 
Hence, the secondary quantized formalism relevant to the field $\sigma (x)$ has to be built up in the space with indefinite metric.

The commutator for dilaton field  is (see (\ref{e24}))
\begin{equation}
\label{e25}
[\bar\sigma (x), \bar\sigma (0) ] = 2\,\pi\,i\,sign (x^{0}) \left [b_{1} \,\theta (x^{2}) + b_{2}\,\delta (x^{2})\right ]. 
\end{equation}
The coefficients $b_{1}$ and $b_{2}$ in (\ref{e25}) can be fixed from the canonical commutation relations 
\begin{equation}
\label{e26}
 \left [A_{\mu}(x), \pi_{A_{\nu}} (0)\right ]_{\vert x^{0} = 0} = i\,g_{\mu\nu}\,\delta^{3}(\vec x)
\end{equation}
and 
\begin{equation}
\label{e27}
 \left [\bar\sigma (x), \pi_{\bar\sigma} (0)\right ]_{\vert x^{0} = 0} = i\,\delta^{3}(\vec x),
\end{equation}
respectively. Here, $\pi_{\bar\sigma} (x)$ and $\pi_{A_{\mu}} (x)$  are the conjugate momenta of $\bar\sigma (x)$ and $A_{\mu} (x)$, respectively.

In order to find the coefficients $b_{1}$ and $b_{2}$ we choose  $I_{\mu}$ in the form 
$I_{\mu} = \epsilon\,m^{2} (\partial_{\mu}\sigma - A_{\mu})$ which is invariant under  gauge transformations (\ref{e11}). 
Hence, we have the new term in LD (\ref{e14}), namely $- (1/2)\,\epsilon\,F_{\mu\nu}\,B^{\mu\nu} + \epsilon\,m^{2} (A_{\mu} - \partial_{\mu}\sigma) (B_{\mu} - \partial_{\mu}\sigma )$ which is Stueckelberg-like [19] Lagrangian, where the $\epsilon$- mixing effect between vector fields $A_{\mu}$ and $B_{\mu}$ is included, and one of the field, $B_{\mu}$, has the mass $m$. There are new equations of motion 
\begin{equation}
\label{e28}
\Box\,A_{\mu} - a\,\partial_{\mu}\,(\partial A) +m^{2} (A_{\mu} - \epsilon^{-1}\,B_{\mu}) = m^{2}\,\partial_{\mu}\sigma,\,\,\, a = 1 - \bar\xi,
\end{equation}
\begin{equation}
\label{e29}
(\Box + \tilde m^{2} )(\partial B) = \tilde m^{2} \Box\sigma,\,\,\, \tilde m^{2} = {\bar\xi^{-1}}\,m^{2} 
\end{equation}
arising instead of Eqs. (\ref{e18}) and (\ref{e17}), respectively. The solution $B_{\mu} = \partial_{\mu}\sigma$  obtained for an arbitrary vector $I_{\mu}$ obeys Eq. (\ref{e29}) that leads to (\ref{e21}).
The solution of Eq. (\ref{e28}) is explicitly given in the form 
\begin{equation}
\label{e30}
 A_{\mu} = C_{\mu} + \frac{1}{m}\,\partial_{\mu}\varphi  - \frac{\bar\xi}{m^{3}}\,\partial_{\mu}\Box\varphi,
\end{equation}
where $\varphi = (1 + \epsilon^{-1})\,m\,\sigma$ with $(\Box +m^{2})C_{\mu} = 0$, $\partial_{\mu} C^{\mu} = 0$ and $[C_{\mu} (x),\varphi (y) ] = 0$. 

Using (\ref{e30}) in (\ref{e26}) with  $\pi_{A_{\nu}} = \epsilon (\partial_{0} B_{\nu} - \partial_{\nu} B_{0} )- \xi\,g_{0\nu}\,\partial^{\rho} B_{\rho}$, one can find 
 $$b_{1} = \frac{1}{(2\,\pi)^{2}}\,\frac{f^{2}}{\bar\xi\, (1 + \epsilon)},$$
while (\ref{e27}) with 
$$\pi_{\bar\sigma} = \left (1 + \frac{\epsilon\,m^{2}}{f^{2}}\right)\,\partial_{0}\bar\sigma - \frac{\epsilon\,m^{2}}{f}\,A_{0} $$ gives 
 $$b_{2} = \frac{-1}{2\,\pi^{2}\,\left (1 - m^{2}/f^{2}\right )}.$$
The propagator of the dilaton field $\bar\sigma (x)$ is
\begin{equation}
\label{e33}
 \tau (x) =  \langle\Omega, T [\bar\sigma (x) \bar\sigma (0)] \Omega\rangle 
  = - b_{1}\left [\ln\vert \kappa^{2} x^{2}\vert + i\pi\theta (x^{2})\right ] +b_{2} 
\left [\frac {1}{x^{2}} + i\pi\delta (x^{2})\right ] + b_{3}, 
\end{equation}
where $\kappa \sim l^{-1}$. On the other hand, (\ref{e33}) can be obtained through the Fourier transformed distribution 
\begin{equation}
\label{e34}
\tau (x)  = \frac{(2\pi)^{2}}{i} \int d_{4}p e^{-ipx}\left \{4b_{1}\lim_{\iota^{2}\rightarrow 0}\left [\frac{1}{(p^{2} -\iota^{2} + i\varepsilon)^{2}} +i\pi^{2}\ln\frac{\iota^{2}}{\kappa^{2}}\,\delta_{4}(p)\right ] + \frac{b_{2}}{p^{2} + i\varepsilon}\right \}.
\end{equation}
To get (\ref{e34}) we have used the following properties of generalized functions [20]
$$\int d_{4}p\,e^{-i\,p\,x} \,\frac{1}{(p^{2} - \iota^{2} + i\,\varepsilon)^{2}} = \frac{i}{8\,\pi^{2}} K_{0}\left (\iota\,\sqrt{-x^{2}_{\mu} + i\,\varepsilon}\right ) $$
for small argument of the Bessel function $K_{0}(z)$
\begin{equation}
\label{e36}
K_{0}(z)\simeq \ln\left (\frac{2}{z}\right) - \gamma + O(z^{2}, z^{2}\,\ln z).
\end{equation}
The coefficient $b_{3}$ in (\ref{e33}) is fixed in such a way as to cancel the term proportional to $\ln 2 - \gamma$ in (\ref{e36}). The second term in the propagator (\ref{e34}) 
breaks the UV stability that means
$$ (-p^{2})^{2}\,\lim_{\iota^{2}\rightarrow 0} \left [\frac{1}{(p^{2} -\iota^{2} + i\varepsilon)^{2}} +i\pi^{2}\ln\frac{\iota^{2}}{\kappa^{2}}\,\delta_{4}(p)\right ] = 1,$$
$$ \frac{1}{(4\,\pi)^{2}}\,\Box^{2} \, \left \{\ln\vert \kappa^{2}\,x_{\mu}^{2}\vert + i\,\pi\,\theta (x^{2}) \right \} = \delta^{4} (p).$$

The commutator of $B_{\mu}$ field is 
$[B_{\mu} (x), B_{\nu} (y) ] = \left (1 + 1/\epsilon\right )^{-2}\,[A_{\mu} (x), A_{\nu} (y)],$
where 
$[A_{\mu} (x), A_{\nu} (0)] = 2\,i/(\pi\,\xi)\,\left (1 + 1/\epsilon\right )\,g_{\mu\nu} \,\varepsilon (x^{0})\,\delta (x^{2}).$
The TPWF $\omega^{B}_{\mu\nu} (x)$ for $B_{\mu}$ field is
$$  \omega^{B}_{\mu\nu} (x) = \frac{g_{\mu\nu}}{\pi^{2}\,\bar\xi (1 + \epsilon)} \frac{1}{x^{2}_{\mu} - i\,\varepsilon\,x^{0}}. $$
The propagator of $B_{\mu}$ field in four-momentum space is 
$\tilde\tau_{\mu\nu} (p) = p_{\mu}\,p_{\nu}\,\tilde\tau (p)$, where
$\tilde\tau (p)  \sim [ \tilde\tau_{1} (p) + \tilde\tau_{2} (p) ] $,
with
\begin{equation}
\label{e42}
\tilde\tau_{1}  (p) = \frac{1}{\bar\xi\,(1 + \epsilon)}\lim_{\iota^{2} \rightarrow 0} \left [\frac{1}{(p^{2} -\iota^{2} +i\,\varepsilon)^{2}} + i\,\pi^{2}\,\ln (l^{2}\,\iota^{2})\,\delta_{4}(p)\right ] ,
\end{equation}
\begin{equation}
\label{e43}
\tilde\tau_{2} (p) = \frac{-1}{2(f^{2} - m^{2})\,(p^{2}  +i\,\varepsilon)},
\end{equation}
where the "strong gauge condition"  $\partial_{\mu}\sigma (x) = B_{\mu} (x)$ was used. 
In case of no mixing between the real photon and DP,
$\tilde\tau_{\mu\nu} (p)$ comes to the standard photon propagator.

{\bf V. Physical and virtual states.-} In the previous section we found the states with an indefinite metric which can not be clarified in terms of physical particles. One can provide these states would not appear in the asymptotic expressions for physical observables at $ t \rightarrow \pm\infty$. For this we follow after Heisenberg [21] and Bogolyubov [22] in $S$-matrix theory. Let us consider the set $\Phi (x)$ of local scalar fields in the form 
$$\Phi (x) = h(x) + \sum_{n} c_{n}\,\chi_{n} (x), \,\,\, c_{n} = const,$$
where some of fields $\chi_{n}(x)\in \{\chi_{1} (x), ..., \bar\sigma (x), ...\}$ may have the commutation relations with negative sign, e.g., (\ref{e25}); $h (x)$ is the set of real (physical) states, e.g., Higgs-boson among them; the dilaton field $\bar\sigma (x)$ stands as a virtual (fictitious) state. We introduce  two Hilbert spaces: the real space $\ S (\Re^{4})$ with respect to real (physical) particles describing by $h(x)$, and  the space $\ S^{\prime} (\Re^{4})$ for $\chi_{n} (x)$ fields. The total Hilbert space is $\mho (\Re ^{4}) =  S (\Re^{4}) +  S^{\prime} (\Re^{4})$. 

We suppose that each amplitude of state possesses by both physical and virtual parts, however the part of the amplitude corresponding to $\chi$ state is defined unique by its physical part based on the state $h$. 
In the operator form $\Phi$ is divided into two parts $\Phi = h + \chi$, where $h = P\,\Phi$ ($h \in S(\Re ^{4})$) and $\chi = (1 - P)\,\Phi$ ($\chi \in S^{\prime}(\Re ^{4})$). Here, $P$ is the operator which projects the states $\Phi$ from $\mho (\Re^{4})$ to $S (\Re ^{4})$; $P ^{+} = P$, $ P^2 = P$; ${\parallel\Phi\parallel}^2 = {\parallel h \parallel}^2 + {\parallel\chi\parallel}^2$, ${\parallel h \parallel}^2 > 0$. In the $S$-matrix approach $\Phi_{+ \infty} = S \,\Phi_{- \infty}$. 
We suppose that the system is found in some state $\Phi_{-\infty} = h_{-\infty} + \chi_{-\infty}$ at $t\rightarrow - \infty$ first, and because of interactions, the system turns into the state $\Phi_{+\infty} = h_{+\infty} + \chi_{+\infty}$ at $t\rightarrow + \infty$. The following condition $\parallel\Phi_{-\infty}\parallel = \parallel\Phi_{+\infty}\parallel$ is evident.
 Then one finds 
\begin{equation}
\label{e44}
h_{+ \infty} = P\,S (h_{- \infty} + \chi_{- \infty}),
\end{equation}
\begin{equation}
\label{e45}
\{ \chi_{- \infty} + (1 - P)\,S (h_{- \infty} + \chi _{-\infty})\} = 0,
\end{equation}
where the nonlocal boundary condition $\chi _{- \infty} + e^{i\,\zeta}\,\chi _{+ \infty} = 0$ is used ($\zeta$ is a phase). The latter does allow to exclude the virtual fields from local equation. 
Eq. (\ref{e45}) defines the unphysical (virtual) part of the amplitude at $t\rightarrow -\infty$ by the physical part of the ampllitude.
$$\chi_{-\infty} = - \{1 + (1 - P)\,S\}^{-1}\, (1 - P)\,S\, h_{-\infty}.$$
The input physical states of the system is described by the vector states of the form:
$$\Phi_{-\infty} = h_{-\infty}  - \{1 + (1 - P)\,S\}^{-1}\, (1 - P)\,S\, h_{-\infty}.$$
From (\ref{e44}) and (\ref{e45}) one finds the asymptotic virtual  $\chi _{+\infty}$ through the asymptotic state $\ h_{-\infty}$ of real particles
\begin{equation}
\label{e46}
 \chi_{+ \infty} = \{ 1 + (1 - P)\,S \}^{-1} (1- P)\,S\, h_{- \infty}.
\end{equation}
Here,  $h_{- \infty}$ is defined through the equation $h_{+\infty} = \tilde S\, h_{-\infty}$ and the unitary matrix $\tilde S$ is 
$\tilde S = PS \{ 1 + (1 - P) S\}^{-1}$. Thus, we may deal with $\tilde S$ as to those $S$-matrix which connects the physical components of $h$ amplitudes of the state only.

{\bf VI. Conclusions.-}  The model for the DP particle solvable in 4-dimensional space-time is studied at the lowest order of perturbative theory using canonical quantization. The model is gauge and scale invariant and these symmetries are spontaneously broken with the following properties: the DP  field is massive.
The interaction between DP and quarks  is mediated by the divergence of the dilaton.
The latter is a portal to the propagator of DP obeying the equation of motion $\Box (\partial\cdot B) =0 $, $\bar\xi \neq 0$. The theories with indefinite metric may lead to Green's functions which are rather regular in UV (see (\ref{e42}), (\ref{e43})). 
The mass of the dilaton field $\bar\sigma$ does not enter the final solutions. This is the consequence the dilaton is virtual (fictitious) state.
We obtained that the states with indefinite metric do not appear in the asymptotic expressions for physical observables.
 The particles in such theories are "transparent" in high energy collisions, but they have the "dark" (or "hidden")  nature.
We estimated the upper limit for the mixing strength $\epsilon < 3 \cdot 10^{-2}$ 
in $S\rightarrow \gamma\bar\gamma$, where the main contribution is due to top-quark in the loop. This can be interpreted as the limit of the branching ratio  $BR (S\rightarrow\gamma\,\bar\gamma)$ which is just the rate of the two-photon decay of the Higgs boson in the SM as $\epsilon = 0$. 
We find that the DP mass $m$ is restricted by 3.3 GeV from above, and the results with EM$\nu$F calculations gives  $m = 0.83$ GeV and $\epsilon = 7.6\cdot 10^{-3}$. 
The decay mode $S\rightarrow\gamma\,\bar\gamma$ can be used to probe the DP sector since the emitted energy of the single photon is encoded with measuring of the missing of the recoil DP.
It's clear from (\ref{e8}) and (\ref{e9}) that the LHC is the most promising machine where DP physics can be discovered.

\end{document}